\documentclass[aps,prd,onecolumn,showpacs,amssymb]{revtex4}
\usepackage{graphicx,epsfig} 
\usepackage{dcolumn}
\usepackage{bm}
\usepackage{caption}
\usepackage{subfig}
\usepackage{amssymb}
\usepackage{color}
\usepackage{amsmath}

\begin{document} 
        
        \title{Evolution of spherical over-densities in tachyon scalar field model}
        \author{M. R. Setare }\email{rezakord@ipm.ir} 
        \affiliation{Department of Science, Campus of Bijar, University of Kurdistan, Bijar , Iran.}
        \author{F. Felegary}\email{falegari@azaruniv.ac.ir}
        \author{F. Darabi }\email{f.darabi@azaruniv.ac.ir; Corresponding
        -author}

        \affiliation{Department of Physics, Azarbaijan Shahid Madani University, Tabriz, 53714-161 Iran}
        
        \date{\today}

\begin{abstract}
We study the tachyon scalar field model in flat FRW cosmology with the particular potential $\phi^{-2}$ and the scale
factor behavior  $a(t)=t^n$.
We consider the spherical collapse model and investigate
the effects of the tachyon scalar field on the structure formation in flat FRW universe. We calculate $\delta_{c}(z_{c})$, $\lambda(z_{c})$, $\xi(z_{c})$, $\Delta_{V}(z_{c})$, $\log [\nu f(\nu)]$ and $\log [n(k)]$ for the tachyon scalar field model  and compare the results with the results of EdS model and $\Lambda CDM$ model. It is shown that in the tachyon scalar field model
the structure formation may occur earlier, in comparison to the other models.
\end{abstract}
\vspace{1cm}
\maketitle

\section{Introduction}

The last  cosmological and astrophysical data of Large Scale structure, the observations 
of type Ia and Cosmic Microwave Background radiation have demonstrated that
currently there is an acceleration expansion phase in the universe \citep{Riess,
bernardis}.
The cosmic expansion can be well described by a negative pressure so-called dark energy (DE). The simplest candidate for DE is the cosmological constant.
However, the cosmological constant suffers from  the fine-tuning and the cosmic coincidence
problems \citep{copeland},\citep{weinberg}.
Therefore, to avoid these problems,  different models for dark energy have been proposed  such as quintessence, K-essence, tachyon \citep{setare}, ghost \cite{malekjani}, phantom, quintom \citep{msetare},  and the quantum gravity models, as well as holographic \cite{holog} and new agegraphic  models \citep{copeland}, \citep{padmanabhan1}. The tachyon model as a scalar field model  arises in particle physics and  string
theory. Thus, it can be considered as one of the potential candidates to describe the nature of the DE. 

On the other hand, the problem of structure formation in the universe is a very important issue
in theoretical cosmology. A simple model of structure formation is
the spherical collapse model.
The spherical collapse model was presented by Gunn and Gott \citep{gun}.
  This model studies the evolution of growth of overdense structures
with respect to the dynamics of scale factor or cosmic redshift. The dynamics of overdense structures depends on the dynamics of the background Hubble flow and expansion of the universe.
In the frame of general relativity, the spherical collapse model has been
studied \citep{fill},\citep{haffman},\citep{ryden}.
In this paper, we study the spherical collapse and the evolution of spherical overdensities in the framework of tachyon scalar field model and compare the results with the results of Einstein-de Sitter (EdS) and $\Lambda$-Cold
Dark Matter ($\Lambda CDM$) models.

\section{Cosmology with Tachyon scalar field }
The Lagrangian of tachyon scalar field over a cosmological background is given by \citep{padd}
\begin{equation}
\mathcal{L}=- V(\phi)\sqrt{1-\partial_{a}\phi~\partial^{a}\phi},
\end{equation}
where  $\phi$ and $V(\phi)$ are the tachyon scalar field and tachyon potential, respectively, and we consider the Friedmann-Robertson-Walker (FRW) metric having the cosmic time $t$ dependent scale factor $a(t)$. For a homogeneous field, the equation of motion is obtained as  
\begin{equation}
\frac{\ddot{\phi}}{1-\dot{\phi}^{2}}+3H\dot{\phi}+\frac{\acute V(\phi)}{V(\phi)}=0,
\end{equation}
where the symbols $^{.}$ and $^\prime$ denote the derivatives with respect to $t$ and $\phi$, respectively, and $H={\dot{a}}/{a}$ is called the Hubble parameter. In the flat FRW universe, the energy density $\rho_{\Lambda}$ and the pressure $p_{\Lambda}$ of the tachyon field read as
\begin{equation}
\rho_{\Lambda}=\frac{V(\phi)}{\sqrt{1-\dot{\phi}^{2}}},\label{rho}
\end{equation}
\begin{equation}
p_{\Lambda}=-V(\phi)\sqrt{1-\dot{\phi}^{2}}.\label{p}
\end{equation}
For the pressureless matter and tachyon scalar field matter, the Friedmann equation is given by 
\begin{equation}
H^{2}=\frac{1}{3M_{pl}^{2}}(\rho_{m}+\rho_{\Lambda}),\label{friedman}
\end{equation}
where $\rho_{\Lambda}$ is  the energy density of tachyon scalar field, and
$\rho_{m}$
is the density of pressureless matter. We suppose that there is no interaction between $\rho_{\Lambda}$ and $\rho_{m}$, so the continuity equations are given separately by
\begin{equation}
\dot{\rho}_{\Lambda}+3H\rho_{\Lambda}(1+\omega_{\Lambda})=0,\label{dot}
\end{equation}
\begin{equation}
\dot{\rho}_{m}+3H\rho_{m}=0.\label{dotm}
\end{equation}
Using Eqs. (\ref{rho}) and (\ref{p}) and also  $p_{\Lambda}=\omega_{\Lambda}\rho_{\Lambda}$, the equation of state  parameter (EoS) for tachyon  scalar field is obtained as
\begin{equation}
\omega_{\Lambda}=\dot{\phi}^{2}-1.\label{eos}
\end{equation}
The requirement for a real $\rho_{\Lambda}$ results in $0<\dot{\phi}^{2}<1$
according to which $\omega_{\Lambda}$ should vary
as $-1<\omega_{\Lambda}<0$. The fractional energy densities
are defined by
\begin{equation}
\Omega_{\Lambda}=\frac{\rho_{\Lambda}}{3M_{pl}^{2}H^{2}},\label{Omega}
\end{equation}
\begin{equation}
\Omega_{m}=\frac{\rho_{m}}{3M_{pl}^{2}H^{2}}.\label{Omega5}
\end{equation}
Taking time derivative of Eq. (\ref{Omega}) and using  Eq. (\ref{dot}) yields
\begin{equation}
\dot{\Omega_{\Lambda}}=-\Omega_{\Lambda}H\Big[3(1+\omega_{\Lambda})+2\frac{\dot{H}}{H^{2}}\Big].\label{domega}
\end{equation}
Also, taking time derivative of Eq. (\ref{friedman}) and using  Eqs. (\ref{dot})
and (\ref{dotm}) yields
\begin{equation}
2\frac{\dot{H}}{H^{2}}=-3(1+\omega_{\Lambda}\Omega_{\Lambda}).\label{Hdot}
\end{equation}
Using Eq. (\ref{Hdot}) and inserting Eq. (\ref{domega}), we obtain
\begin{equation}
\acute \Omega_{\Lambda}=3\omega_{\Lambda}\Omega_{\Lambda}(\Omega_{\Lambda}-1).
\end{equation}
Here, the prime is the derivative with respect to $x=\ln a$ where $a=(1+z)^{-1}$
and z is the cosmic redshift. Using $\frac{d}{dx}=-(1+z)\frac{d}{dz}$
and Eq. (\ref{eos}), one finds
\begin{equation}
\frac{d\Omega_{\Lambda}}{dz}=-3\Omega_{\Lambda}(\Omega_{\Lambda}-1)(\dot{\phi}^{2}-1)(1+z)^{-1}.\label{df}
\end{equation}
The differential equation for the evolution of dimensionless Hubble
parameter, $E(z)=\frac{H}{H_{0}}$, in tachyon scalar field model, is obtained
by using Eqs. (\ref{dot}), (\ref{dotm}), (\ref{eos}) and (\ref{Hdot}) as
follows 
\begin{equation}
\frac{dE}{dz}=\frac{3}{2}\frac{E}{(1+z)}\Big[1+\Omega_{\Lambda}(\dot{\phi}^{2}-1)\Big].\label{ddf}
\end{equation}
Now, we consider the following particular potential which results in the scalar field with linear time dependence  and   the scale factor with suitable
power law behaviour,
 as follows \cite{padd}
\begin{equation}
V(\phi)=\frac{2n}{M_{pl}^{2}}(1-\frac{2}{3n})^{\frac{1}{2}}\frac{1}{\phi^{2}},
\end{equation}
\begin{equation}
\phi=\sqrt{\frac{2}{3n}}t,\label{sqr}
\end{equation}
\begin{equation}
a(t)=t^{n}.\label{scale}
\end{equation}
Taking time derivative of Eq. (\ref{sqr}), inserting in Eq. (\ref{eos}) and using Eq. (\ref{scale}), we can obtain the equation of state parameter for tachyon scalar field model
\begin{equation}
\omega_{\Lambda}=\frac{2}{3n}-1.\label{omegalam}
\end{equation}
In Eq. (\ref{omegalam}), we can see that if $n\geq\frac{2}{3}$, then we will have $-1<\omega_{\Lambda}<0$.
Using Eq. (\ref{omegalam}) and inserting it in Eqs. (\ref{df}), (\ref{ddf}) we can get the evolution of  EoS parameter $(\omega_{\Lambda})$, the
density parameter of dark energy $(\Omega_{\Lambda})$, and the dimensionless Hubble parameter $(E(z))$
in tachyon scalar field model as a function of cosmic redshift. In Fig.1, assuming the present values $\Omega_{\Lambda_{0}}\approx 0.7$,  $\Omega_{m_{0}}\approx0.3$ and $H_{0}\approx67.8 \frac{km}{s Mpc}$,
we have shown the evolution of EoS parameter, the  evolution of
density parameter and  the evolution of dimensionless Hubble parameter of tachyon scalar field model with respect to the redshift parameter $z$ for the typical value $n=1$.

\begin{figure}
\centering

{\includegraphics[width=2in]{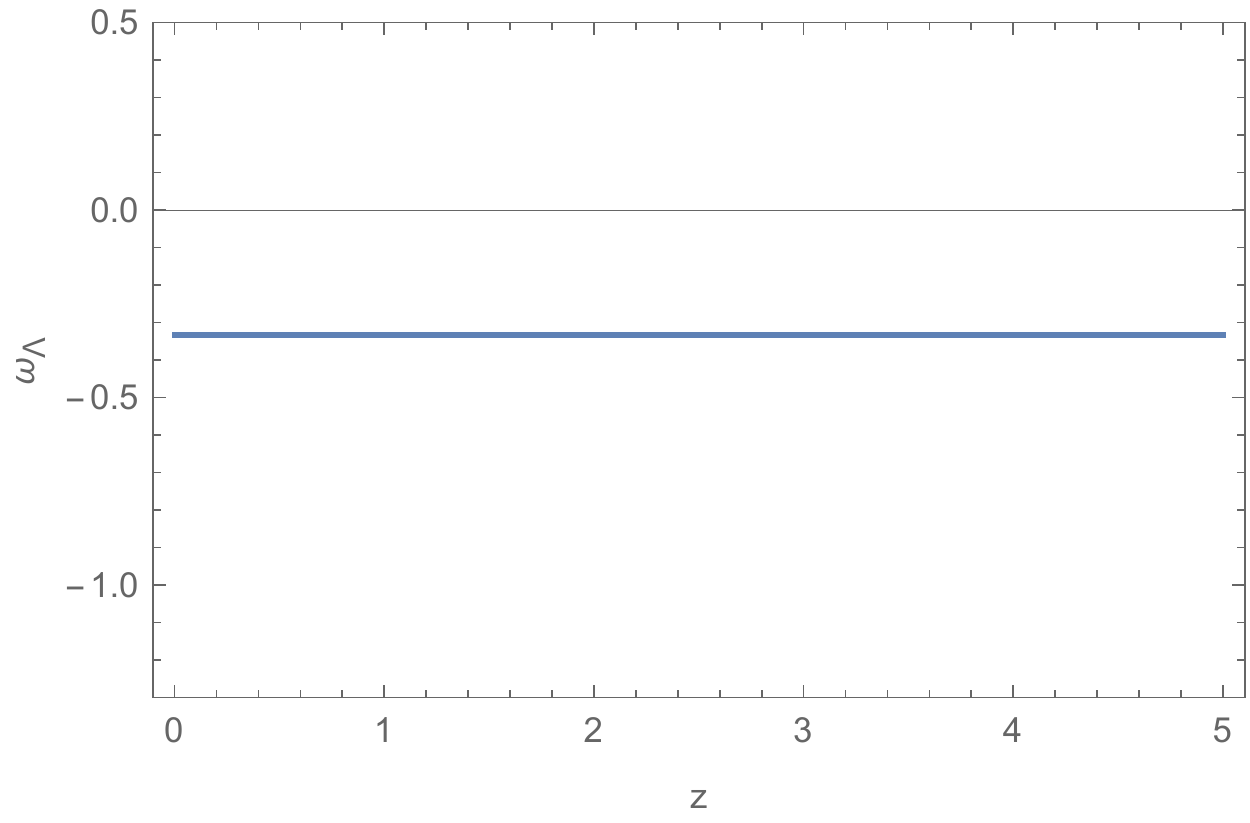}}

{\includegraphics[width=2in]{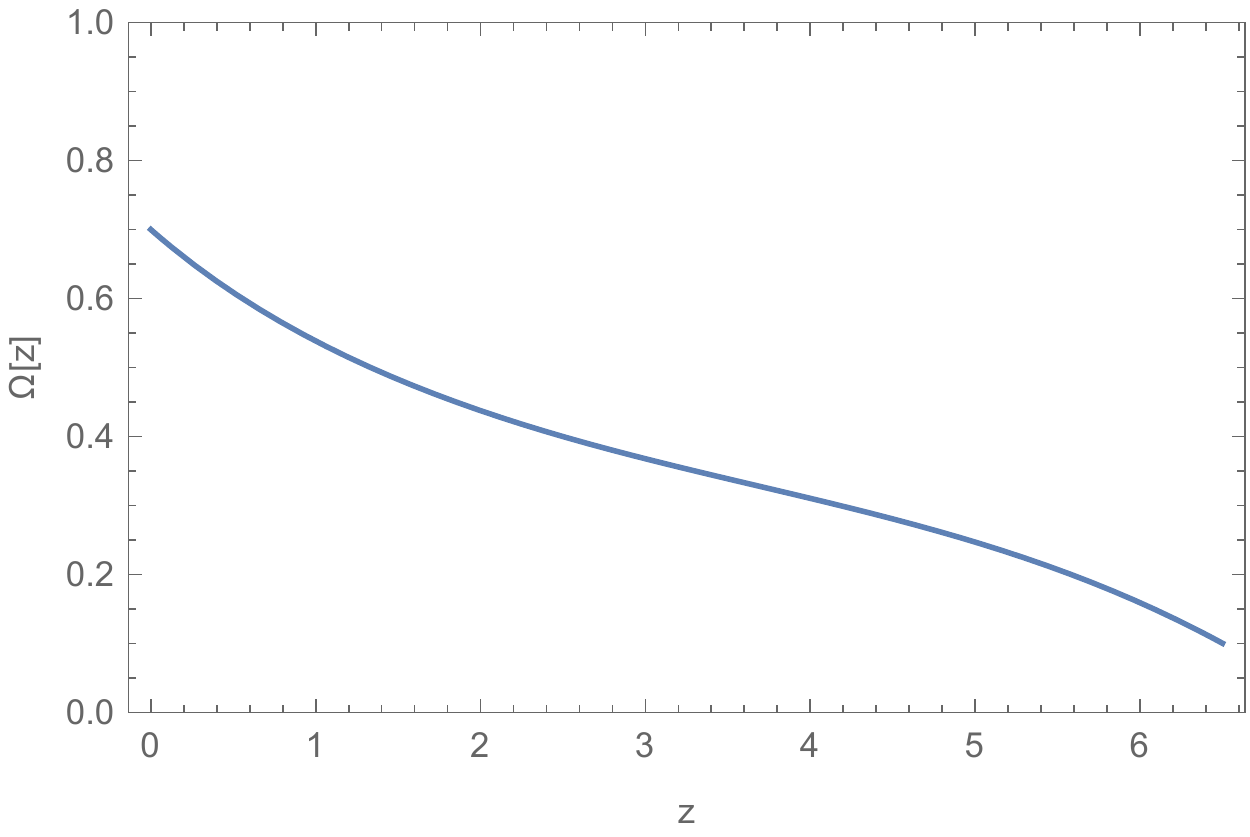}}

{\includegraphics[width=2in]{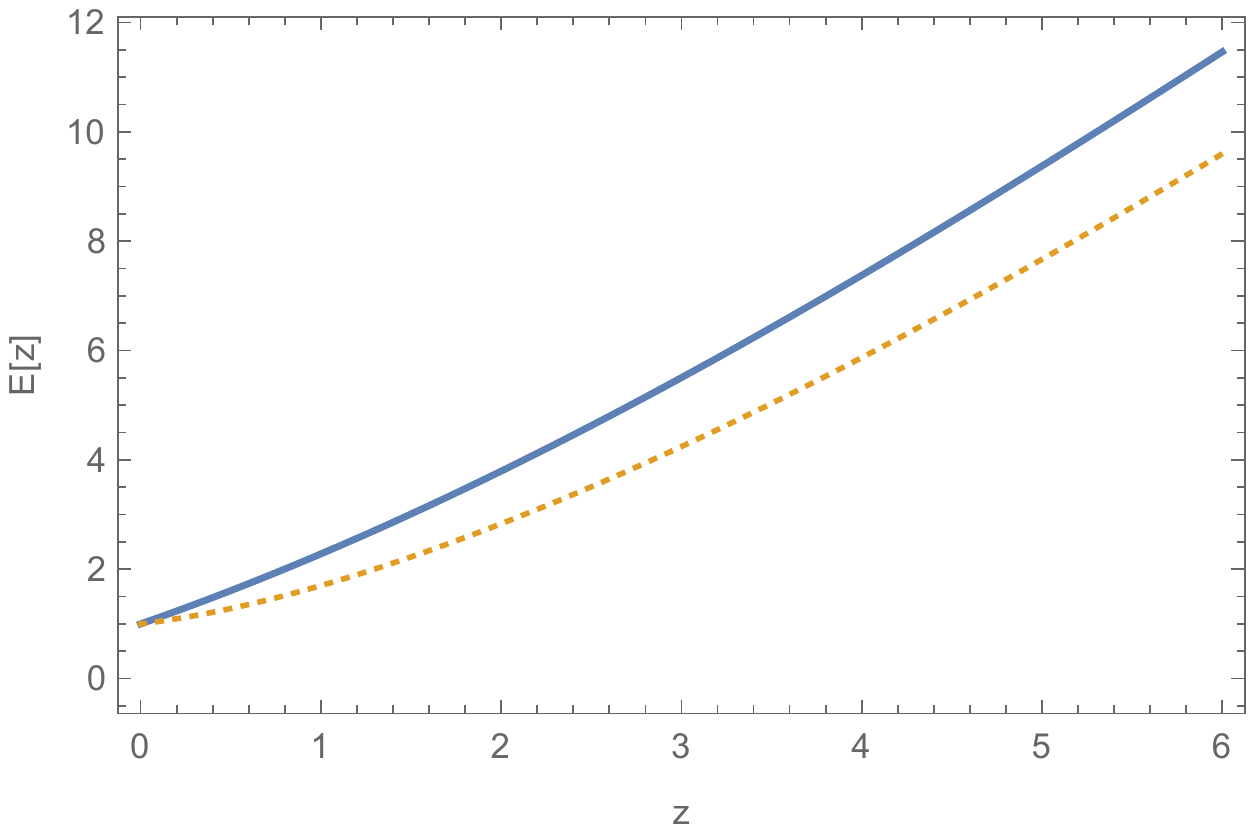}}

\caption{{The evolution of EoS parameter (top), dark energy density parameter
(middle),
 and dimensionless Hubble parameter (down) of tachyon scalar field model
 with respect to the redshift parameter $z$.} The thick line represents the tachyon scalar field model for $n=1$ and the dotted line shows the $\Lambda CDM$ model.}
\end{figure}

\section{linear perturbation theory}
In this section, we study the linear growth of perturbation of non relativistic dust matter by computing the evolution of growth factor $g(a)$ in tachyon scalar field
model, and then compare it with the evolution of growth factor in EdS and
$\Lambda CDM$ models. The differential equation for 
the evolution of the growth factor $g(a)$ is given by \citep{pace},\citep{pace1},\citep{percival}
(see Appendix I)

\begin{equation}
g^{''}(a)+(\frac{3}{a}+\frac{E^{'}(a)}{E(a)})g^{'}(a)-\frac{3}{2}\frac{\Omega_{m_{0}}}{a^5E^2(a)}g(a)=0.\label{ga}
\end{equation} 
In order to study the linear growth in tachyon scalar field
model, using Eqs.(\ref{df}), (\ref{ddf}) and (\ref{omegalam}) for $n=1$,  we solve numerically Eq.(\ref{ga}). 
To obtain the linear growth of structures in the EdS model and the $\Lambda CDM$ model, we use the procedure used in Ref \cite{holog}.
\begin{figure} 
  \centering
  \includegraphics[width=2in]{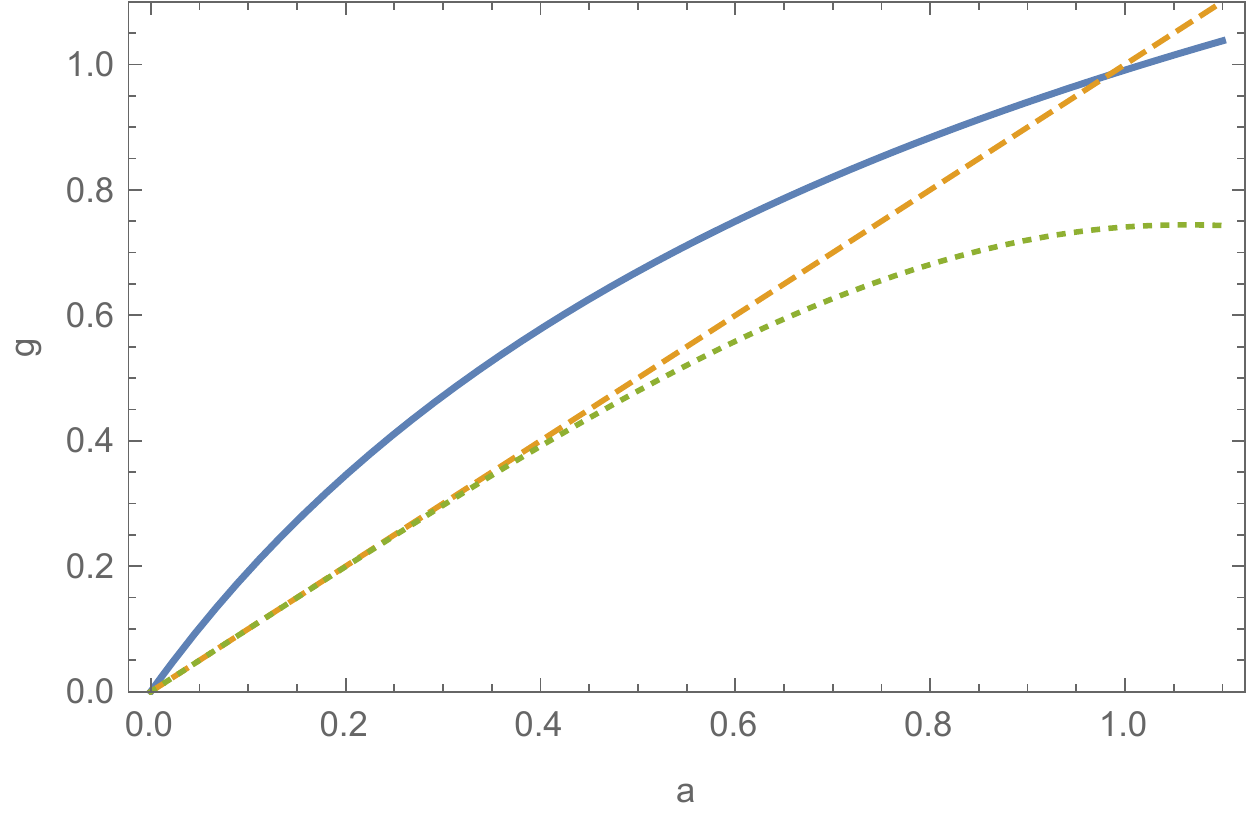}~
    \caption{Time evolution of the growth factor as a function of the scale
    factor. The thick line represents the tachyon scalar field model for  $n=1$. The dotted line indicates the  $\Lambda CDM$ model and  the dashed 
  line represent the $EdS$ model.}
\end {figure}
In figure (2), we have plotted the evolution of growth factor $g(a)$ with respect to  the scale factor. At first, namely for small scale factors, the growth factor
in the tachyon scalar field model is larger than those of EdS and
 $\Lambda CDM$ models.  However, for rather larger scale factors,  the growth factor in the tachyon scalar field model becomes smaller than the EdS model while it is still larger enough than that of $\Lambda CDM$ model. This means
that, at the beginning, the tachyon scalar field model
predicts structure formation more efficient than EdS and
 $\Lambda CDM$ models. For later times, however, the structure formation in the tachyon scalar field model is dropped behind  that of EdS model, whereas it precedes the structure formation in the  $\Lambda CDM$ model.

\section{spherical collapse in the tachyon scalar field model}
 The structure  formation   is described by a non-linear differential
equation for the evolution of the matter perturbation $\delta$  in a matter
dominated universe \citep{bernard},\citep{padmanabhan}. In \citep{abramo} this
differential equation was generalized
to  the case of evolution of $\delta$ in a universe including
a dark energy component. Now, we consider the non-linear differential equation
which is given by \citep{pace} (see Appendix I)
\begin{equation}
\delta^{''}+\left(\frac{3}{a}+\frac{E^{'}(a)}{E(a)}\right)\delta^{'}-\frac{4}{3}\frac{{\delta^{'}}^{2}}{1+\delta}
-\frac{3}{2}\frac{\Omega_{m_{0}}}{a^5E^2(a)}\delta(1+\delta)=0,\label{delta}
\end{equation} 
where $'$ denotes the derivative with respect to $a$. In the linear regime,
we have
\begin{equation}
\delta^{''}+\left(\frac{3}{a}+\frac{E^{'}(a)}{E(a)}\right)\delta^{'}
-\frac{3}{2}\frac{\Omega_{m_{0}}}{a^5E^2(a)}\delta=0.\label{delta1}
\end{equation} 
\begin{figure}
\centering
\subfloat[][The linear growth of density perturbation $\delta$ in terms of
$a$.]
{\includegraphics[scale=0.4]{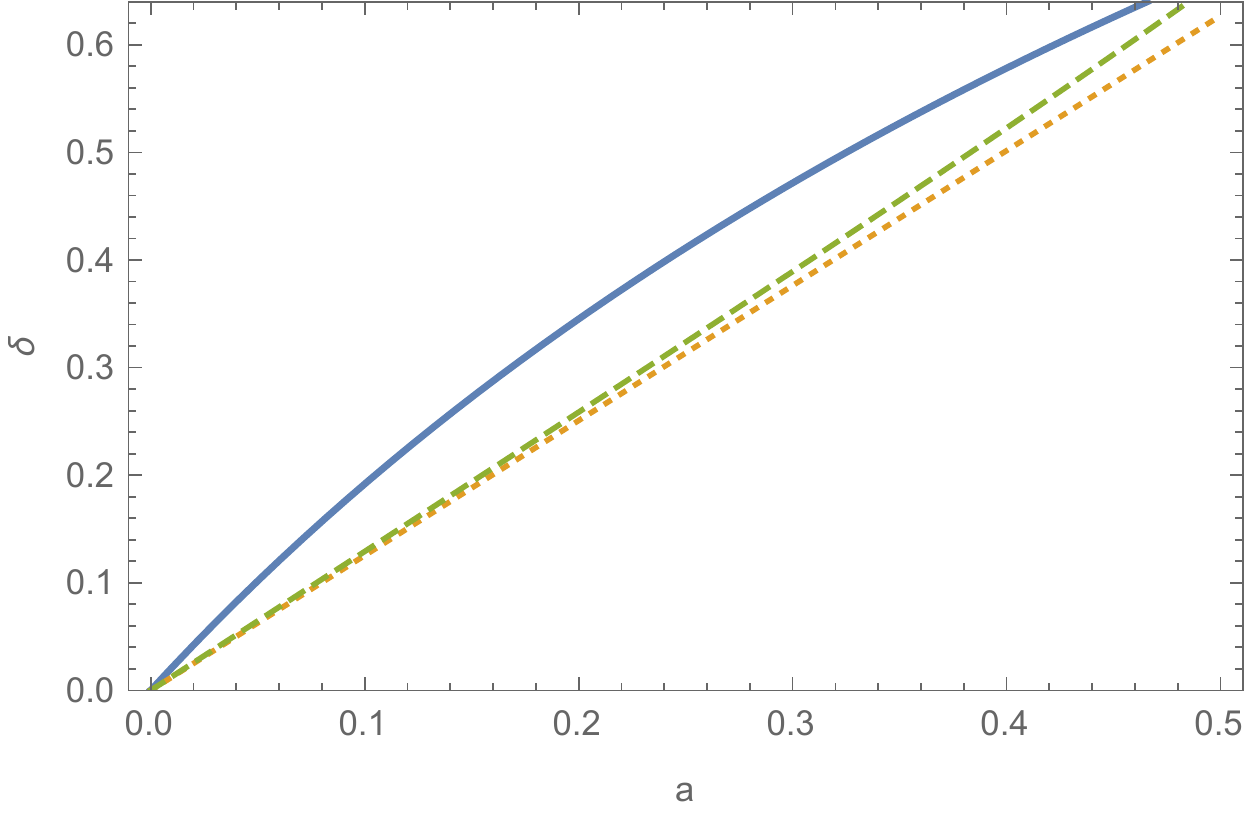}}~~~~~~~~
\subfloat[][The non-linear growth of density perturbation $\delta$ in terms of $a$.]
{\includegraphics[scale=0.4]{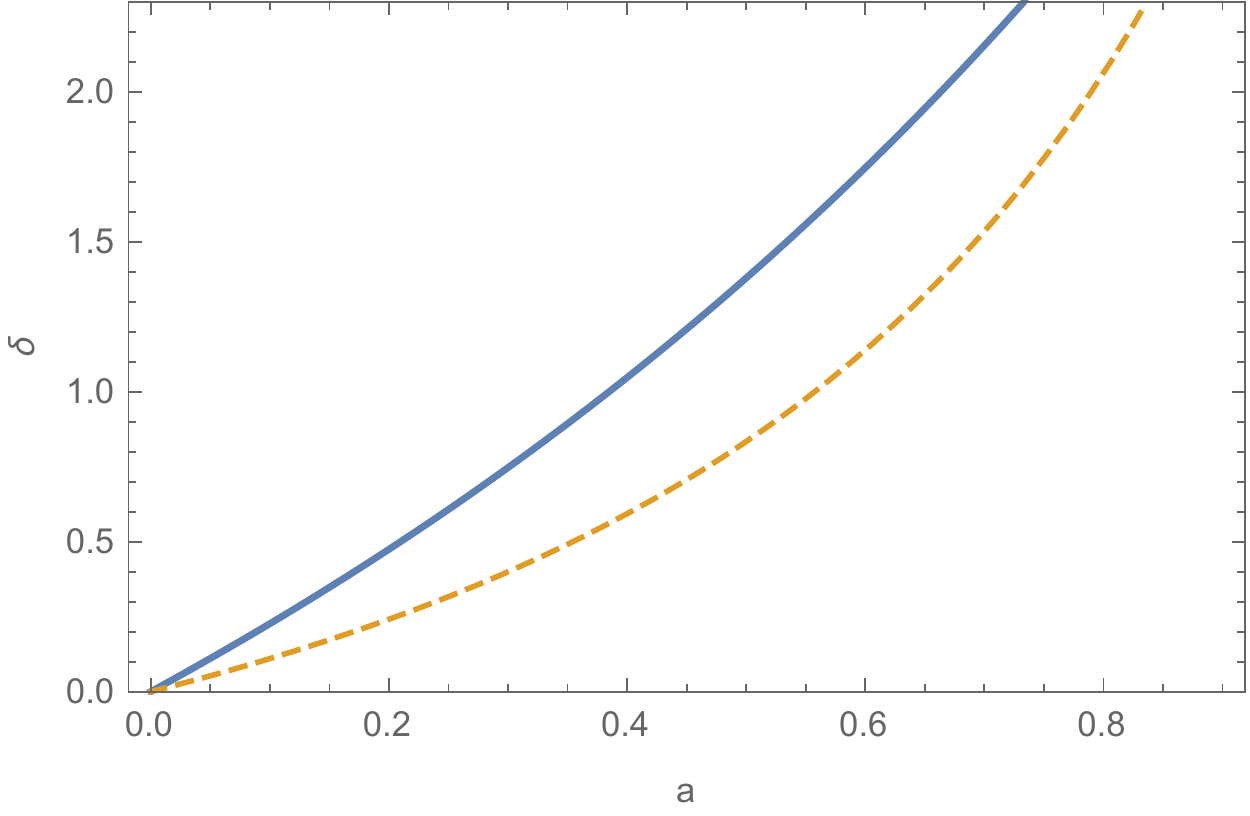}}~~~~~~~~
\caption{The thick line represents the tachyon scalar field model for $n=1$. The dotted line indicates the $\Lambda CDM$ model and the dashed line indicates the EdS model.}
\end{figure}

 In EdS model,  we consider the initial conditions $\delta_{i}=20.9\times10^{-4}$,
$\acute \delta_{i}=0$ and $a_{i}=10^{-4}$  \citep{pace}. In order to study the linear growth of density perturbation and the non-linear growth of density perturbation in tachyon scalar field
model, using Eqs.(\ref{df}), (\ref{ddf}) and (\ref{omegalam}) for $n=1$,  we solve numerically Eqs.(\ref{delta}) and (\ref{delta1}) (see Appendix
 II).
To obtain the linear growth of density perturbation $\delta$ in the EdS model and the $\Lambda CDM$ model, we use the procedure used in Ref \cite{holog}.
The figure (3-a)  shows that  the linear growth factor in the tachyon scalar field model is larger than those of EdS and the $\Lambda CDM$ models, and
the figure  (3-b) shows that the non-linear growth factor in the tachyon scalar field model is larger than the EdS model.

\section{Determination of $\delta_c$  and $\Delta_{V}$} 
As time passes, the perturbation is growing and one can no longer use the linear regime. At this stage, the radius   of perturbation region becomes maximal $R=R_{max}$ and the perturbation stops  growing. This  condition is called turn-around which points to the epoch when the grows
of perturbation decouples from the Hubble flow of the homogenous background. After the turn-around the perturbation starts contracting. For a perfect
spherical symmetry and perfect pressureless matter, the perturbation would
collapse to a single point becoming infinitely dense. Since there is
hardly any perfect spherical symmetric overdensity in the universe, 
the perturbation does not collapse to a single point and finally a virialized object of a certain finite size
in equilibrium state is formed that is called halo.

We call ($z_{c}$, $R_{c}$) and ($z_{ta}$, $R_{ta}$) as the redshift and radius corresponding to virialization and the turn-around epochs, respectively. Now, we peruse two characterising quantities of the spherical collapse model
for the tachyon scalar field  model: the virial overdensity $\Delta_{V}$ and the linear overdensity parameter  $\delta_c$.
We consider a
spherical overdense region with matter density $\rho$ 
 in a surrounding universe described by its background dynamics and density $\rho_{b}$.
The virial overdensity $\Delta_{V}$ is defined by \citep{meyer}

\begin{equation}
\Delta_{V}=\frac{\rho}{\rho_{b}}\frac{R_{c}}{a_{c}},
\end{equation}
which is a function of scale factor and redshift.  We can rewrite the virial overdenty $\Delta_{V}$ as follows \citep{meyer}

\begin{equation}
\Delta_{V}=1+\delta(a_{c})=\xi(\frac{x_{c}}{\lambda})^{3},
\end{equation}
where
\begin{equation}
x_{c}=\frac{a_{c}}{a_{ta}},
\end{equation}
\begin{equation}
\xi=\frac{\rho(R_{ta})}{\rho_{b}(a_{ta})}=1+\delta(a_{ta}).
\end{equation}
Here, $\lambda$ is the virial radius which is given by \citep{ws}
\begin{equation}
\lambda=\frac{1-\frac{\eta_{\nu}}{2}}{2+\eta_{t}-\frac{3}{2}\eta_{\nu}} ,
\end{equation} 
where $\eta_{t}$ and $\eta_{\upsilon}$  are the (Wang-Steinhardt) WS  parameters 
\begin{equation}
\eta_{t}=\frac{2}{\xi}\frac{\Omega_{\Lambda}(a_{ta})}{\Omega_{m}(a_{ta})},
\end{equation} 
\begin{equation}
\eta_{\nu}=\frac{2}{\xi}\frac{\Omega_{\Lambda}(a_{c})}{\Omega_{m}(a_{c})}\Big(\frac{a_{ta}}{a_{c}}\Big).
\end{equation}

\begin{figure}
  \centering
  \includegraphics[width=2in]{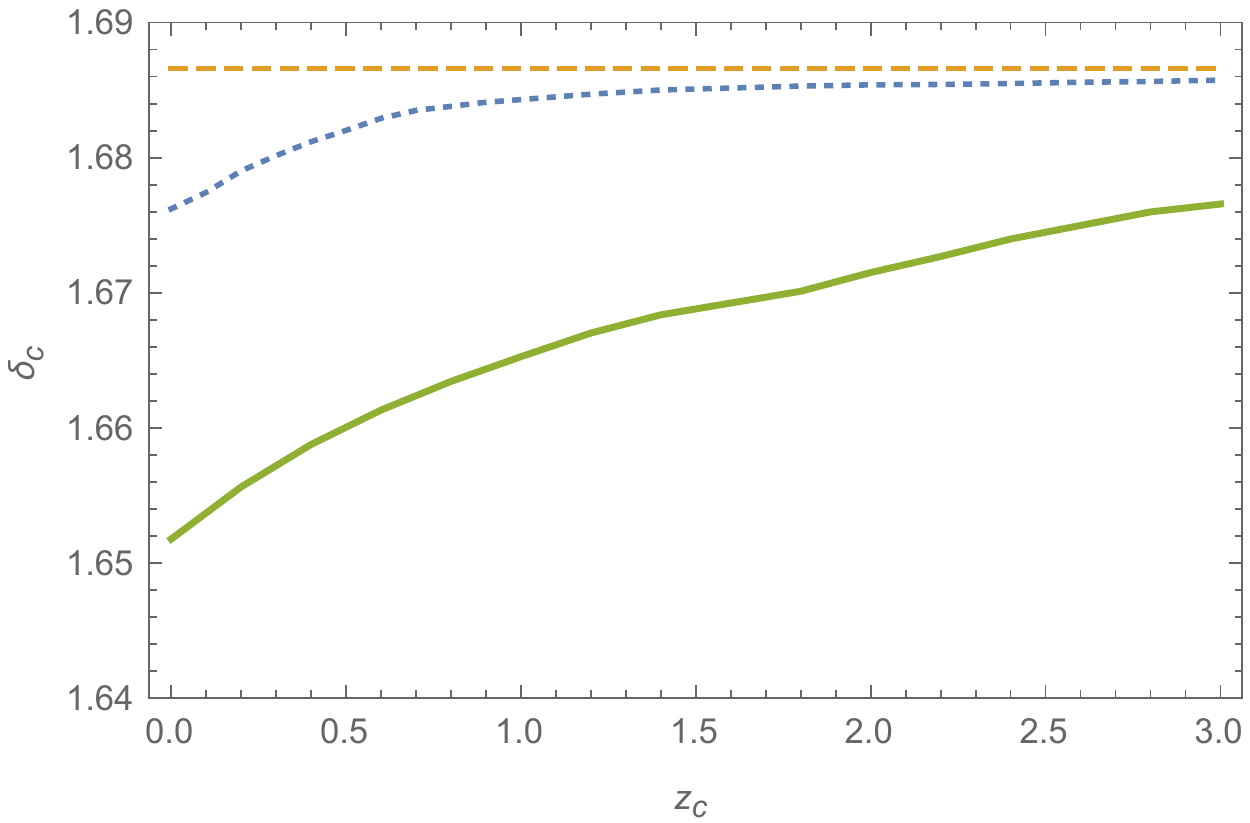}~
    \caption{The time evolution of the linear overdensity, $\delta_{c}(z)$,
in terms of a function of the collapse redshift for the $\Lambda CDM$ model, the EdS
model and the tachyon scalar field model.
The thick line represents the tachyon scalar field model for  $n=1$. The dotted line indicates the  $\Lambda CDM$ model and  the  dashed line indicates the EdS model. }
\label{t2}
\end {figure}

Now, we discuss the results obtained for the linear overdensity parameter and the
virial overdensity for the models introduced in this work.
The figure (\ref{t2}) shows the time evolution of linear overdensity, $\delta_{c}(z)$
in terms of a function of the collapse redshift for the $\Lambda CDM$ model, the EdS
model and the tachyon scalar field model. In the EdS model, $\delta_{c}$ is independent of the redshift, hence it has a constant value i.e. $\delta_{c}=1.686$. In the $\Lambda CDM$ model, $\delta_{c}$ is smaller than 1.686 but the time evolution of the linear overdensity approaches  the value of the EdS model at high redshifts.

In fact, at high redshifts we have a matter dominated  universe (dust matter),
but at lower redshifts we have a dark energy dominated  universe, thus the structure formation  must occur earlier. In the tachyon scalar field model,
$\delta_{c}$ drives more slowly than the $\Lambda CDM$ and the EdS  models
because in Fig.1, the Hubble parameter in  the tachyon scalar field model  is larger than  that of  $\Lambda CDM$ model. 

\begin{figure}
\centering
\includegraphics[width=2in]{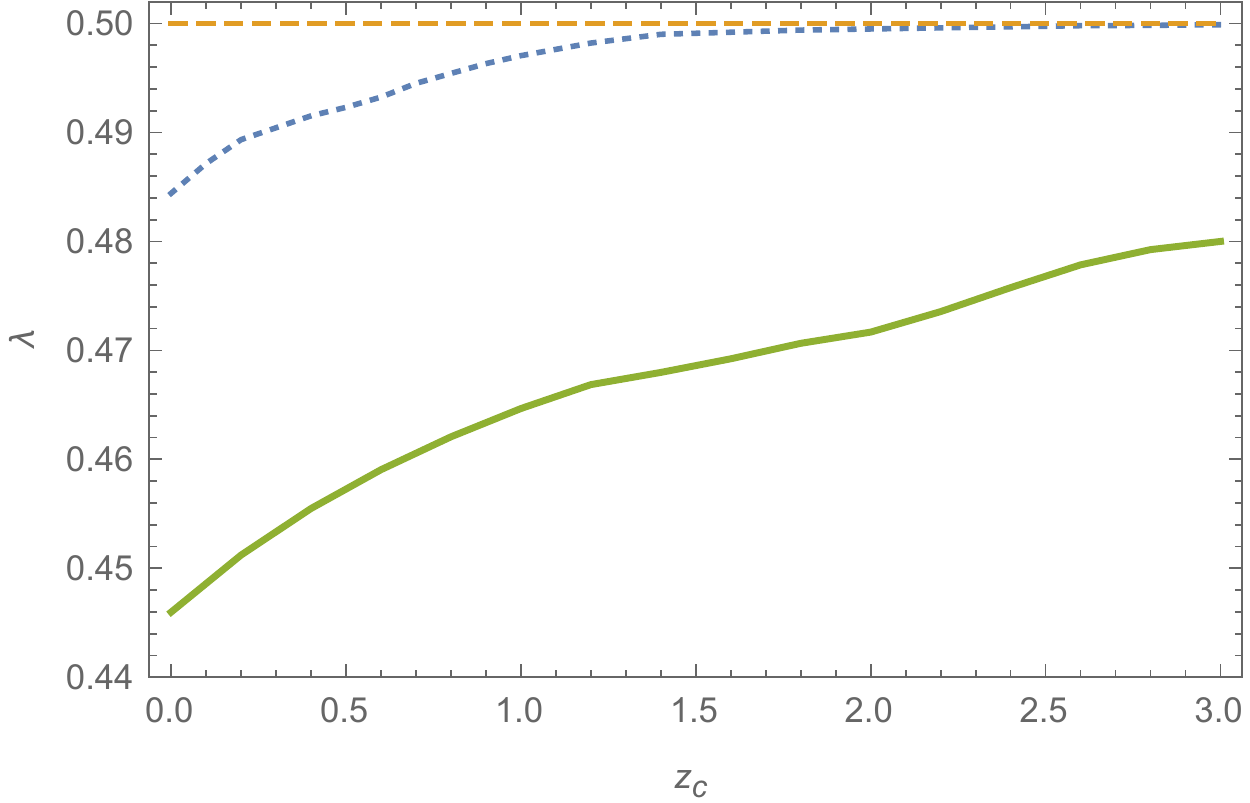}~
\caption{The virial radius $\lambda(z_{c})$ in terms of the collapse redshift
$z_{c}$ for the $\Lambda CDM$ model, the EdS model
and the tachyon scalar field model.
The thick line represents the tachyon scalar field model for  $n=1$. The dotted line indicates the  $\Lambda CDM$ model and  the  dashed line indicates the EdS model.}
\label{u2}
\end {figure} 

In the figure (\ref{u2}), we represent $\lambda(z_{c})$ in terms of $z_{c}$  for the $\Lambda CDM$ model, the EdS model and the tachyon scalar field model.
In the EdS model, $\lambda(z_{c})$ is independent of the redshift, thus it has a constant value i.e. $\lambda=0.5$ . In the $\Lambda CDM$ model, $\lambda(z_{c})$ is smaller than 0.5 but it approaches  the value of the EdS model at high redshifts. In the tachyon scalar field model,
$\lambda(z_{c})$ drives more slowly than the $\Lambda CDM$ and the EdS  models  but its value approaches  the value of the EdS model at high redshifts. Therefore, we can conclude  that the size of structures in the $\Lambda CDM$ model is larger than the tachyon scalar field model.

\begin{figure}
\centering
\subfloat[][The variation of $\xi(z_{c})$ - $z_{c}$ for the $\Lambda CDM$ model, the EdS model and the tachyon scalar field model.]
{\includegraphics[scale=0.4]{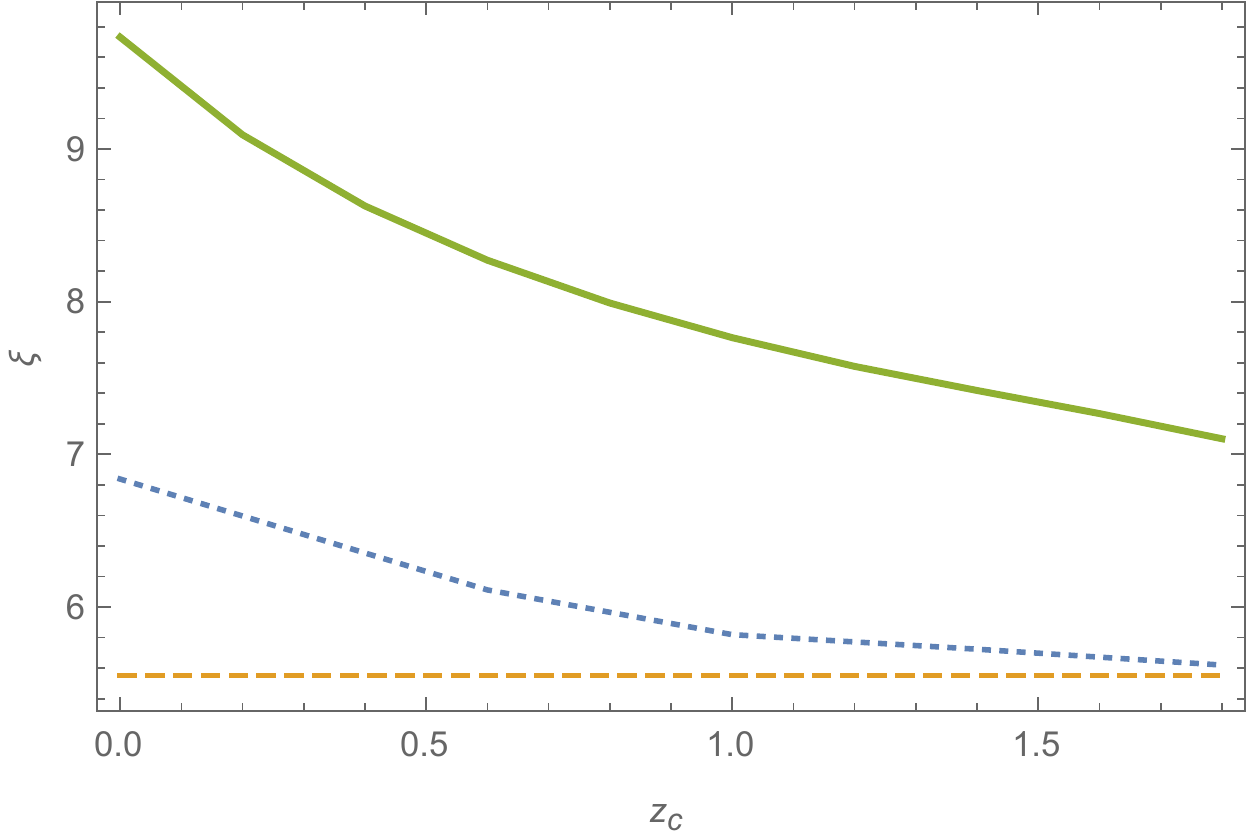}}~~~~~~~~
\subfloat[][The variation of $\Delta_{V}(z_{c})$ - $z_{c}$ for the $\Lambda CDM$ model, the EdS model and the tachyon scalar field model.]
{\includegraphics[scale=0.4]{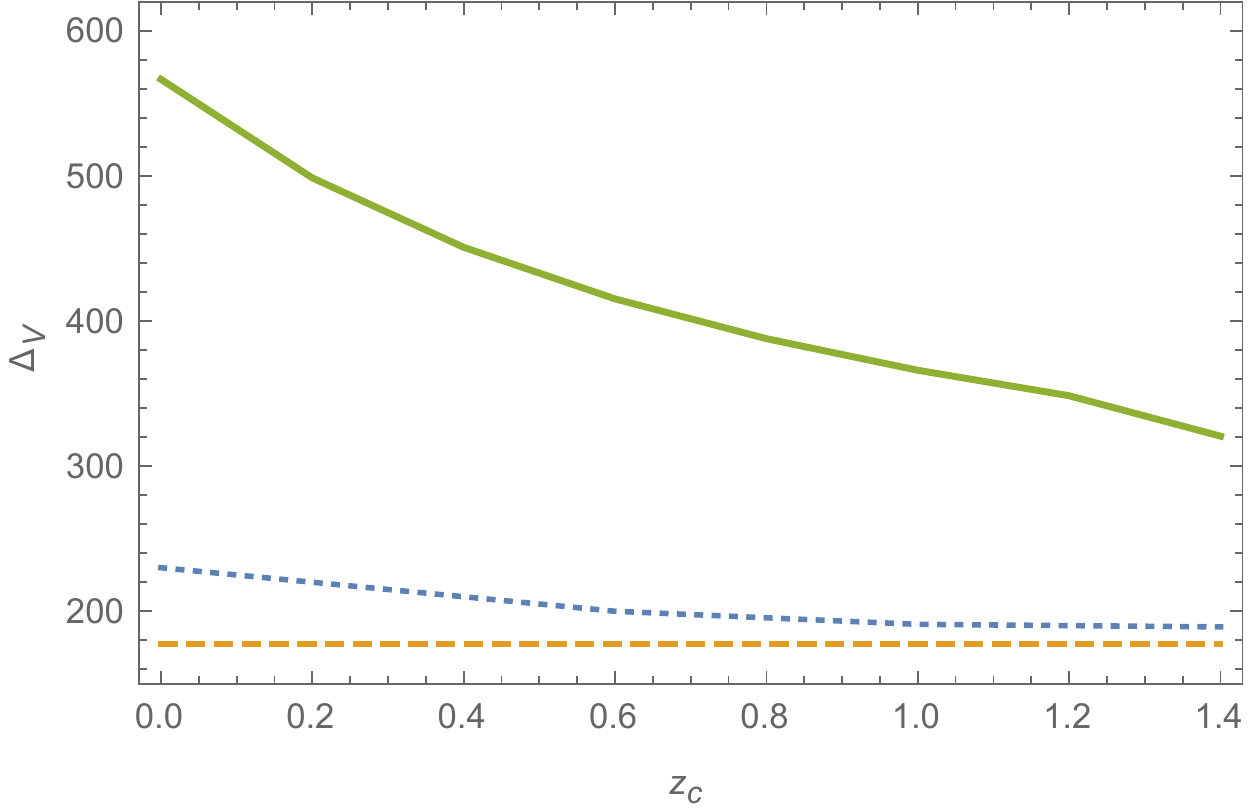}}~~~~~~~~
\caption{The blue, green and red lines represent the tachyon scalar field model for  $n=0.9~,1,~1.1$, respectively. The dotted line indicates the  $\Lambda CDM$ model and  the  dashed line indicates the EdS model. }
\end{figure}

In the figure (6-a),  we represent $\xi(z_{c})$ in terms of $z_{c}$  for the $\Lambda CDM$ model, the EdS model and the tachyon scalar field model.
In the EdS model, $\xi(z_{c})$ is independent of redshift thus it has a constant value i.e. $\xi=5.6$. In the $\Lambda CDM$ model, $\xi(z_{c})$ is larger than 5.6 but its value approaches  the value of the EdS model in terms of high redshifts. In the tachyon scalar field model,
$\xi(z_{c})$ drives faster than the $\Lambda CDM$ model and the EdS  model but its value approaches  the value of the EdS model in terms of high redshifts. Therefore,  we can conclude that in the tachyon scalar field model the overdense spherical regions in terms of $z_{c}$ are denser than the $\Lambda CDM$ model  and the EdS model.

In the figure (6-b), we show  the virial overdensity $\Delta_{V}(z_{c})$ in terms of $z_{c}$  for the $\Lambda CDM$ model, the EdS  model and the tachyon scalar field model.
In the EdS model, $\Delta_{V}(z_{c})$ is independent of redshift, thus it has a constant value, $\Delta_{V}=178$ . In the $\Lambda CDM$ model, $\Delta_{V}(z_{c})$ drives more faster than 178, but its value approaches  the value of the EdS model in terms of high redshifts. In the tachyon scalar field model,
$\Delta_{V}(z_{c})$ drives faster than the $\Lambda CDM$ model  and its
value approaches  the value of the EdS model in terms of high redshifts. The evolution of virial overdensity parameter $\Delta_{V}(z_{c})$  is the main quantity for the halo size. Therefore,
we can conclude that in the tachyon scalar field  model the halo size is
larger than the EdS  model  and the $\Lambda CDM$ model.

\section{Number density and mass function}
The average comoving number density of halos of mass
M is given by \citep{bond},\citep{press}
\begin{equation}
n(M,z)=(\frac{\rho}{M^2})\frac{d \log \upsilon}{d \log M}\nu f(\nu),\label{n}
\end{equation}
where $f(\upsilon)$ and $\rho$ are the multiplicity function and the background density, respectively and  $\nu$ is  given by 
\begin{equation}
\nu=\frac{\delta_{c}^2}{\sigma^2(M)}.
\end{equation}
Here $\sigma(M)$ is the r.m.s of the mass fluctuation in sphere of mass M.
We can use the formula given by
\cite{viana}
\begin{equation}
\sigma(M,z)=\sigma_{8}(z)\Big(\frac{M}{M_{8}}\Big)^{-\frac{\gamma(M)}{3}},\label{1}
\end{equation}
where $\sigma_{8}$ is the mass variance of the overdensity on the scale of
$R_{8}$, $M_{8}=6\times10^{14}\Omega_{m_{0}}h^{-1}M_{\odot}$ and $R_{8}=8h^{-1}Mpc$
are the mass  and the radius inside a sphere.
Also, $\sigma_{8}(z)$ is given by
\begin{equation}
\sigma_{8}(z)=g(z)\sigma_{8}(M,z=0),
\end{equation}
where $g(z)$ is the linear growth factor, $\sigma_{8,DE}(M,z=0)=0.8\times\Big(\frac{\delta_{c,DE}(z=0)}{\delta_{c,\Lambda
CDM}(z=0)}\Big)$ and    
\begin{equation}
\gamma(M)=(0.3\Gamma+0.2)\Big[2.92+\frac{1}{3}\log (\frac{M}{M_{8}})\Big],\label{2}
\end{equation}
where  $\Gamma=\Omega_{m_{0}}h \exp(-\Omega_{b}-\frac{\Omega_{b}}{\Omega_{m_{0}}})$.  Eqs. (\ref{1}) and (\ref{2}) have a validation range \citep{viana}. They
 express that the fitting formula predicts higher values of the
 variance for $M<M_{8}$ and the fitting formula predicts lower values of the variance for $M>M_{8}$. Following  \citep{sherth},
 we apply ST mass function 
 \begin{equation}
\nu f_{ST}(\nu)=0.3222\sqrt{\frac{0.707\nu}{2\pi}}\Big[1+(0.707\nu)^{-0.3}\Big]\exp(-\frac{0.707\nu}{2}).\label{st}
\end{equation}  
We use  the mass function introduced by del Popolo (PO mass function)\citep{del}
\begin{eqnarray}\label{del}
\nu f(\nu)=1.75\sqrt{\frac{0.707\nu}{2\pi}}\Big[1+\frac{0.1218}{(0.707\nu)^{0.585}}+\frac{0.0079}{(0.707\nu)^{0.4}}\Big]
\nonumber \\ \times \exp\Big[-0.4019\times 0.707\nu\Big(1+\frac{0.5526}{(0.707\nu)^{0.585}}+\frac{0.02}{(0.707\nu)^{0.4}}\Big)^{2}\Big],
\end{eqnarray}  
Also, we use the mass function  (YNY mass function) presented in \citep{yahagi}
\begin{equation}
\nu f(\nu)=0.298\Big[1+(0.893\sqrt{\frac{\nu}{2}})^{1.39}\Big]\nu^{(\frac{0.408}{2})}
\exp \Big[-(0.893\sqrt{\frac{\nu}{2}})^{2}\Big].\label{y}
\end{equation}  

Now, we represent the evolution of the ST mass function with respect to $k$ ($k=\log(\frac{M}{M_{8}}))$ in Fig.7  for the tachyon scalar field model and the $\Lambda CDM$ model. We can see that the evolution of  ST mass function with respect to $k$  is the same for tachyon scalar field and the $\Lambda CDM$ models in the  $z=0$ case, but it is different for tachyon scalar field and $\Lambda CDM$ models in the  $z=1$
case.

\begin{figure}
\centering
\subfloat[][The evolution of the mass function with respect to $k$ 
 for tachyon scalar field model and $\Lambda CDM$ model in the case $z=0$.]
{\includegraphics[scale=0.4]{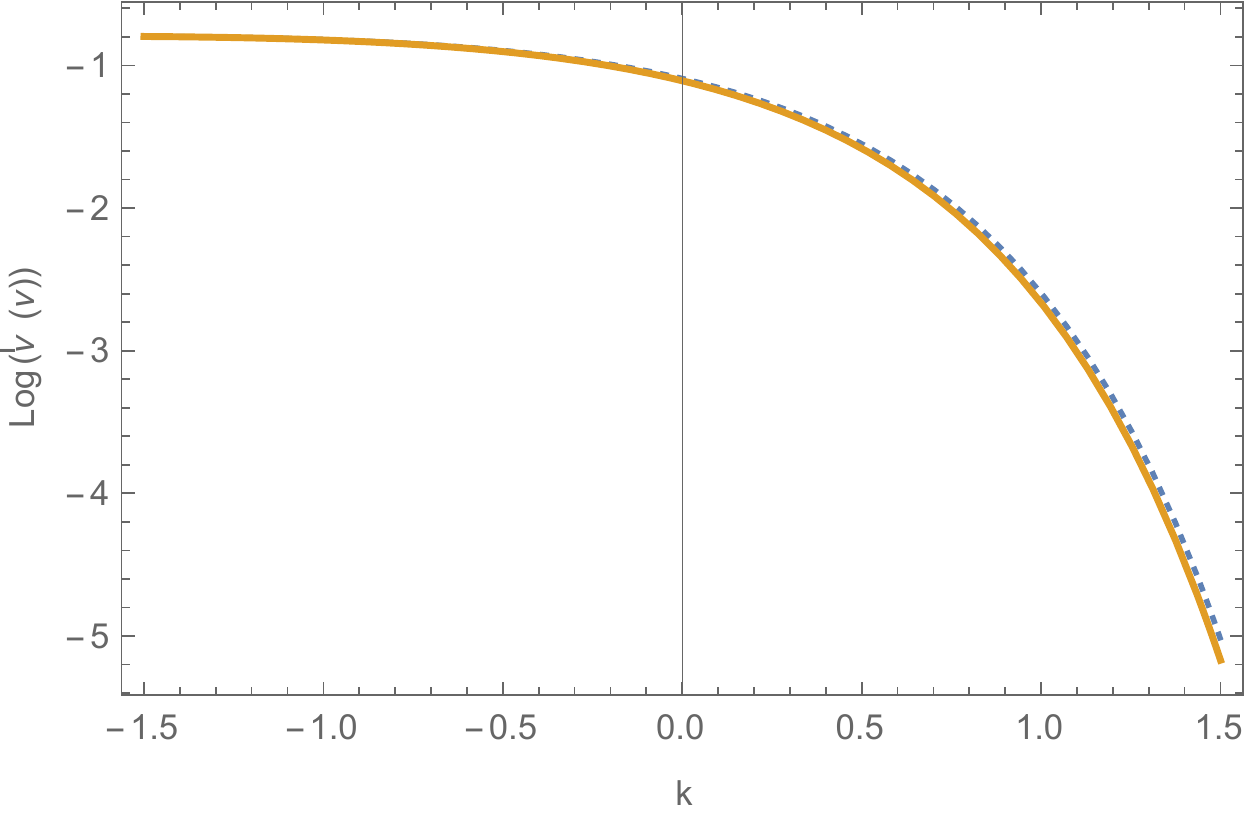}}~~~~~~~~
\subfloat[][The evolution of the mass function with respect to $k$ 
 for tachyon scalar field model and $\Lambda CDM$ model in the case $z=1$.
]{\includegraphics[scale=0.4]{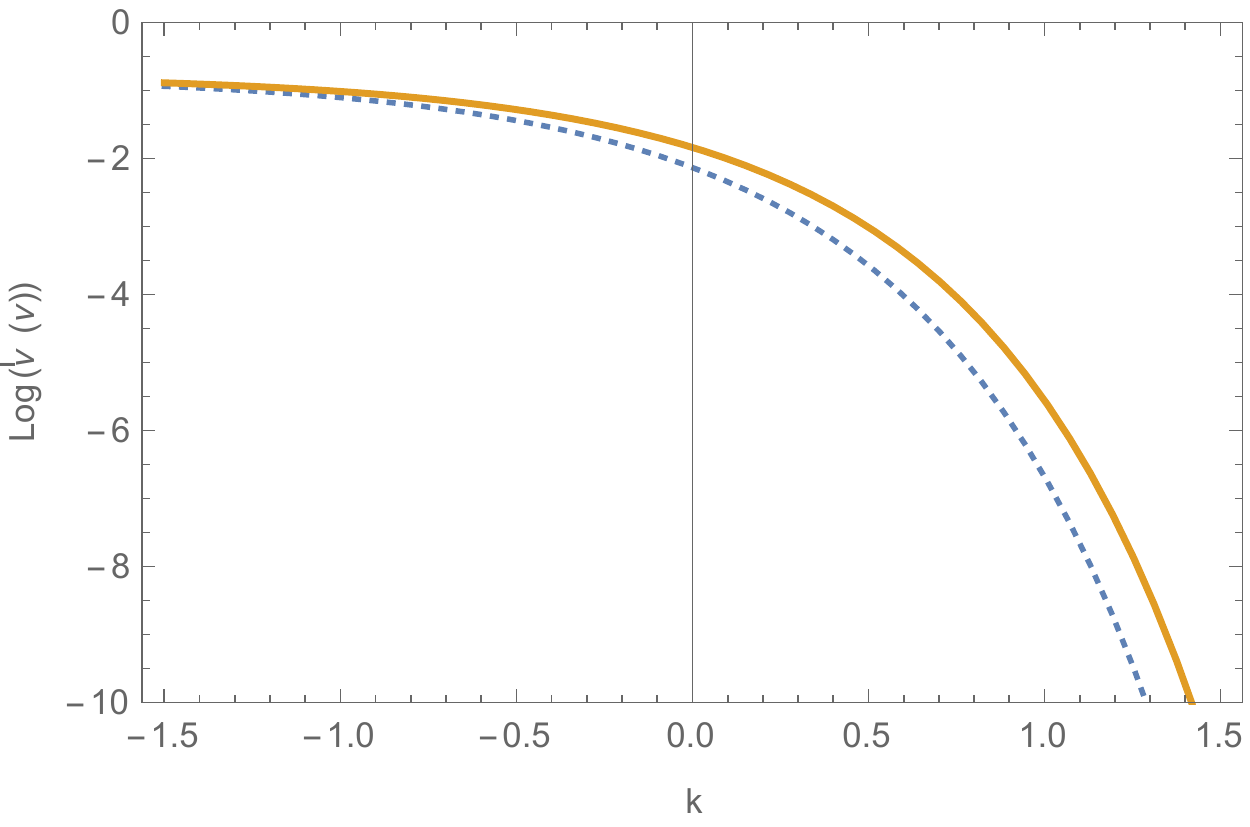}}~~~~~~~~
\caption{The thick line represents the tachyon scalar field model for  $n=1$ and The dotted line indicates the  $\Lambda CDM$ model. }
\label{jop}
\end{figure}

 Using Eqs. (\ref{n})
and (\ref{st}), for the tachyon scalar field model and the $\Lambda CDM$ model, we obtain the average comoving number density of halos of mass $M$  in the cases $z=0~,1$. In Fig.\ref{jop}, we can see  explicitly the differences for the cases $z=0$ and $z=1$.
  We can see that difference of the number densities of halo objects is negligible for small objects in the case  $z=1$. Therefore,  we can obtain  the number density of halo objects for high mass, and we find that the number of objects per unit mass is increasing for high mass in the tachyon  scalar field model.
Also, using Eqs. (\ref{st}), (\ref{del}) and (\ref{y}), we can compare the various mass functions at $k=0$ in Fig.\ref{jop11}. We can see that the PO mass function is larger than ST mass function and YNY mass function for all mass scales.

\begin{figure}
        \centering
        \subfloat[][The evolution the number densities of halo objects
        for the tachyon scalar field model and  the $\Lambda CDM$ model in the case $z=0$.]
        {\includegraphics[scale=0.4]{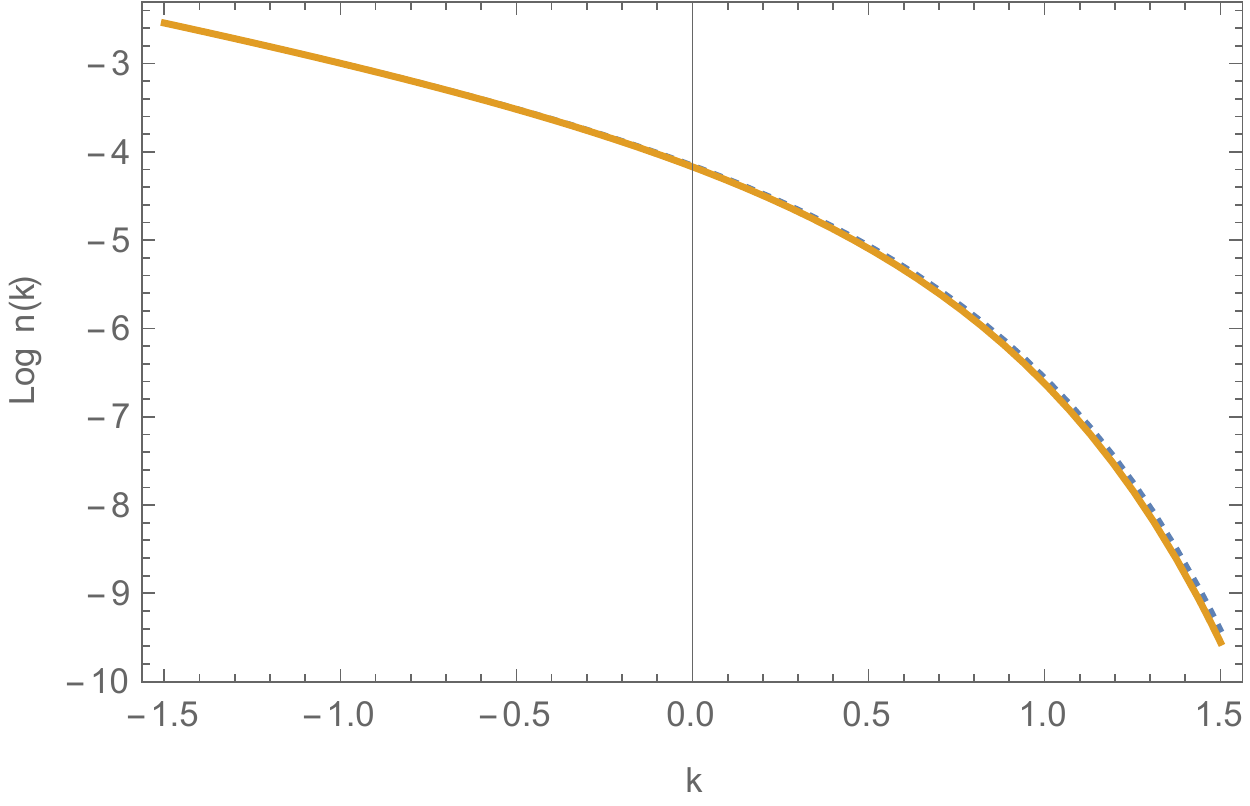}}~~~~~~~~
        \subfloat[][The evolution the number densities of halo objects
        for the tachyon scalar field model and the $\Lambda CDM$ model in the case $z=1$.]
        {\includegraphics[scale=0.4]{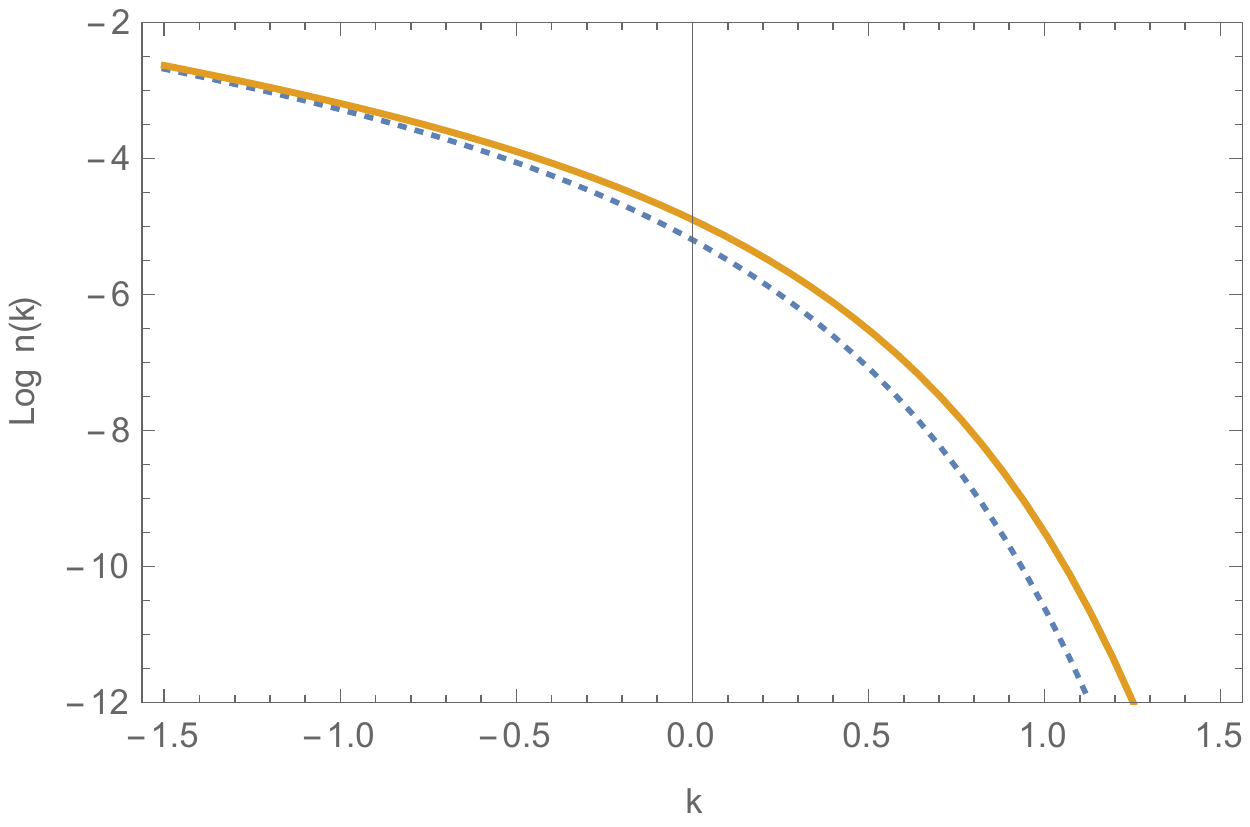}}~~~~~~~~
        \caption{The thick line represents the tachyon scalar field model for  $n=1$ and the dotted line indicates the  $\Lambda CDM$ model. }
        \label{jop}
\end{figure}

\begin{figure}
\centering{\includegraphics[width=2in]{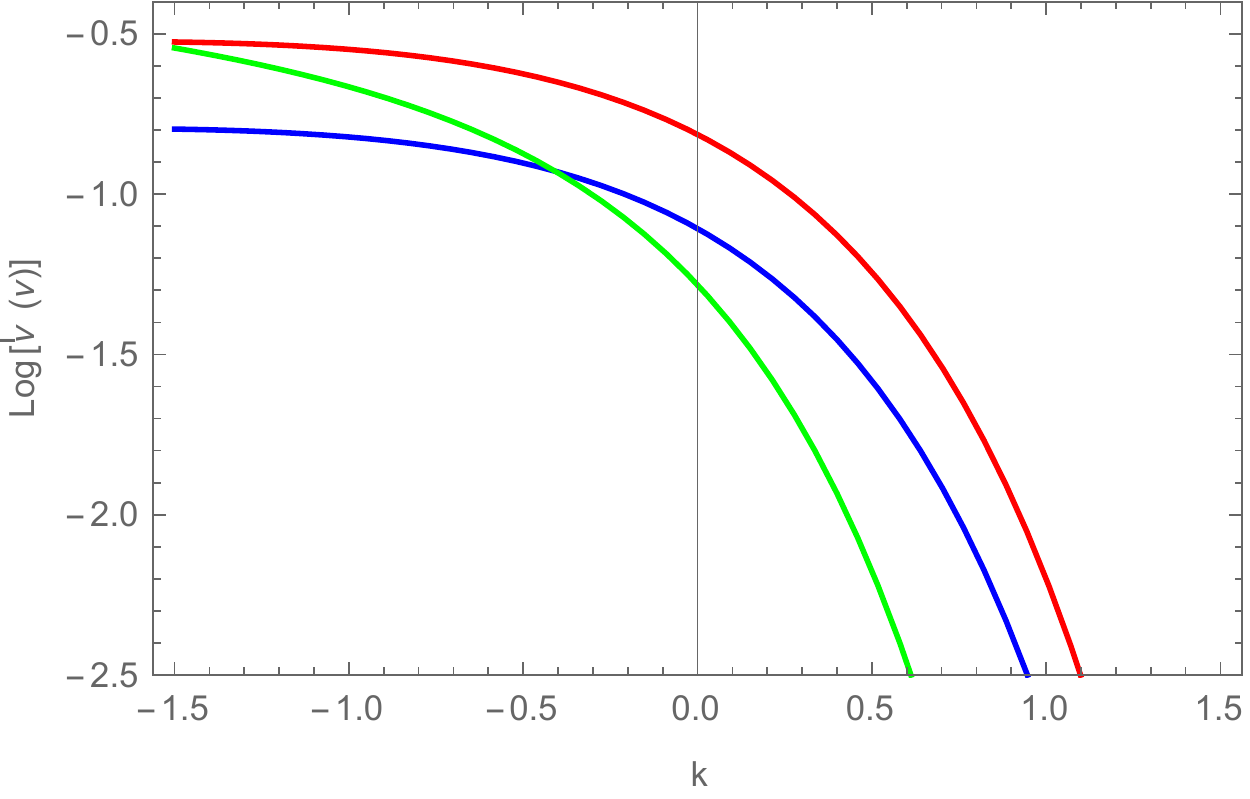}}
\caption{ The evolution of the various mass functions with respect to $k$ for the tachyon scalar field model, $n=1$, in the case
$z=0$. The blue thick line represent ST mass function, the red thick line represent PO mass function and
the green thick line represent YNY mass function }
\label{jop11}
\end{figure}

\section{Conclusion}
In this paper, we  have studied the evolution of spherical overdensities in tachyon scalar field model by assuming a particular potential and the scale factor with power law behavior. We have shown the evolution of the EoS parameter, the  evolution of
the density parameter and  the evolution of the dimensionless Hubble parameter of tachyon scalar field model with respect to a function of $z$, for a typical
value $n=1$.
We have also shown that  at early times of the scale factor evolution, the growth factor
in the tachyon scalar field model drives faster than the EdS 
and $\Lambda CDM$ models. So, it is concluded that in the tachyon scalar field model the structure formation may occur sooner than in the other models. At later times, however, we have shown that the growth factor in the tachyon scalar field model drives more slower than that of EdS model. Also,
in the EdS model, $\delta_{c}$ is independent of the redshift and thus it has a constant value  $\delta_{c}=1.686$.
In the $\Lambda CDM$ model, $\delta_{c}$ is smaller than 1.686, but the time evolution of the linear overdensity approaches the value of  EdS model at high redshifts. In fact, at high redshifts, we have a matter dominated universe (dust matter), but at lower redshifts we have a dark energy dominated  universe, thus the structure formation occurs earlier. In the tachyon scalar field model, $\delta_{c}$ is driven more slower than that of  $\Lambda CDM$ and EdS models, because in figure (1), the Hubble parameter in  the tachyon scalar field model  is larger than  that of $\Lambda CDM$ model. 

Also, we have shown that in the EdS model, $\lambda(z_{c})$ is independent of the redshift, thus it has a constant value i.e. $\lambda=0.5$ . Moreover,
the size of structures in the $\Lambda CDM$ model was larger than that of
 tachyon scalar field model.
In the EdS model, $\xi(z_{c})$ is independent of the redshift, hence it has a constant value  i.e. $\xi=5.6$ . We have shown that in tachyon scalar field model,  the overdense spherical regions with respect to $z_{c}$ are denser than those of $\Lambda CDM$ and EdS models.
In the EdS model, $\Delta_{V}(z_{c})$ is independent of the redshift, so it has a constant value, $\Delta_{V}=178$ . The evolution of virial overdensity parameter $\Delta_{V}(z_{c})$  is the main quantity for the halo size. Therefore,
we  have found that in the tachyon scalar field  model the halo size is
larger than the EdS  and $\Lambda CDM$ models.

Finally, we have shown that the evolution of the ST mass function with respect to $k$  is the same for  tachyon scalar field and $\Lambda CDM$ models in the $z=0$ case, but it is not the same for  tachyon scalar field  model and  the $\Lambda CDM$ model in the $z=1$ case. Also, the evolution of the number density with respect to $k$ is the same for the tachyon scalar field and $\Lambda CDM$ models in the  $z=0$ case, but  its evolution is
not the same for the tachyon scalar field and $\Lambda CDM$ models in the $z=1$ case. The difference of number densities of halo objects is negligible for small objects in the   $z=1$ case. Therefore, in  obtaining  the number density of halo objects for high mass, we find that the number of objects per unit mass is increasing for high mass in the tachyon  scalar field model.

\section*{\textbf{\textit{Appendix I}}}

The Lagrangian of tachyon scalar field over a cosmological background is given by \citep{padd}
\begin{equation}\label{c7}
\mathcal{L}=- V(\phi)\sqrt{1-\partial_{a}\phi~\partial^{a}\phi}~,
\end{equation}
where  $\phi$ and $V(\phi)$ are the tachyon scalar field and tachyon potential, respectively. One can obtain the energy-momentum tensor of the tachyon scalar field as follows \cite{Zhang.}
\begin{equation}\label{c9}
T_{\mu\nu}=\frac{V(\phi) \partial_{\mu}\phi \partial_{\nu}\phi}{\sqrt{1+g^{\alpha \beta}\partial_{\alpha}\phi \partial_{\beta}\phi}}
-g_{\mu \nu} V(\phi) \sqrt{1+g^{\alpha \beta}\partial_{\alpha}\phi \partial_{\beta}\phi}~.
\end{equation}
Now, considering  (\ref{c9}) as describing a perfect fluid, the energy density $\rho$ and the pressure $p$ for the tachyon scalar field are given by 
$$\rho=-T_{0}^{0}=\frac{V(\phi)}{\sqrt{1-\dot{\phi}^{2}}},$$
$$p=T_{i}^{i}=-V(\phi)\sqrt{1-\dot{\phi}^{2}}.$$
 
The fundamental equations for cosmic fluid in Newtonian gravity are defined as follows \cite{pace}
\begin{equation}\label{x2}
\frac{\partial \rho}{\partial t}+\bigtriangledown_{\vec{r}}~.(\rho \vec{v})+\frac{p}{c^{2}}\bigtriangledown _{\vec{r}}~.\vec{v}=0,
\end{equation} 
\begin{equation}\label{x3}
\frac{\partial \vec{v}}{\partial t}+(\vec{v}~.\bigtriangledown _{\vec{r}})\vec{v}+\bigtriangledown _{\vec{r}}\Phi+\frac{c^{2}\bigtriangledown _{\vec{r}}p+\vec{v}\dot{p}}{\rho c^{2}+p}=0,
\end{equation} 
\begin{equation}\label{x4}
\bigtriangledown^{2}\Phi=4 \pi G \Big(\rho+\frac{3p}{c^{2}}\Big),
\end{equation} 
\begin{equation}\label{x5}
\dot{\bar{\rho}}+3H\Big(\bar{\rho}+\frac{p}{c^{2}}\Big)=0,
\end{equation} 
where $\vec{v}$ is the velocity in three-space, $\Phi$ is the Newtonian gravitational potential, $\vec{r}$ is the physical coordinate and $\bar{\rho}$ is 
the density of cosmic  background. 
Now, we use the comoving coordinates as follows \cite{pace}:
\begin{equation}\label{x6}
\vec{r}=a \vec{x}.
\end{equation}
Here $\vec{r}$, $a$ and $\vec{x}$ are the physical coordinates, the scale factor and the comoving coordinates, respectively. Taking time derivative of Eq. (\ref{x6}),   
one can obtain \begin{equation}\label{x7}
\vec{v}(\vec{x},t)=a\Big[H(a)\vec{x}+\vec{u}(\vec{x},t) \Big], 
\end{equation}
where 
\begin{equation}\label{xk7}
\vec{v}(\vec{x},t)=\frac{dr(\vec{x},t)}{dt}, 
\end{equation}
\begin{equation}\label{xk17}
\vec{u}(\vec{x},t)=\frac{dx(\vec{x},t)}{dt}. 
\end{equation}
Here, $H(a)$ is the Hubble function and $\vec{u}(\vec{x},t)$ is the comoving peculiar velocity. Next, one can introduce the following definitions 
\begin{equation}\label{x8}
\bigtriangledown_{\vec{r}}=\frac{1}{a} \bigtriangledown_{\vec{x}}~,
\end{equation} 
\begin{equation}\label{xk8}
\frac{\partial}{\partial t}\Big|_{r}=\frac{\partial}{\partial t}\Big|_{x}-\frac{1}{a}\vec{v}.\bigtriangledown_{\vec{x}}~,
\end{equation}
\begin{equation}\label{x9}
\rho(\vec{x},t)~=~ \bar{\rho}(1+\delta(\vec{x},t))~,
\end{equation}
\begin{equation}\label{x10}
p~=~\omega \rho(\vec{x},t) c^{2}~,
\end{equation}
\begin{equation}\label{x11}
\Phi(\vec{x},t)~=~\Phi_{0}(\vec{x},t)+\phi(\vec{x},t)~,
\end{equation}  
where $\omega$ is the equation of state parameter. Now, using Eqs. (\ref{x7}), (\ref{x8}), (\ref{xk8}), (\ref{x9}), (\ref{x10}), (\ref{x11}) and inserting Eqs. (\ref{x2}), (\ref{x3}), (\ref{x4}) and (\ref{x5}), one can obtain \cite{pace}
\begin{equation}\label{x12}
\dot{\delta}+(1+\omega)(1+\delta)\bigtriangledown_{\vec{x}}~.~\vec{u}=0,
\end{equation}
\begin{equation}\label{x13}
\frac{\partial \vec{u}}{\partial t}+2H \vec{u}+ (\vec{u}~.\bigtriangledown _{\vec{x}})\vec{u}+\frac{1}{a^{2}}\bigtriangledown _{\vec{x}}\phi=0,
\end{equation}
\begin{equation}\label{x14}
\bigtriangledown_{\vec{x}}^{2}\phi-4 \pi G \Big(1+3\omega\Big)a^{2}\bar{\rho}\delta=0.
\end{equation} 
We take the divergence of the Eq. (\ref{x13}) and represent the analysis \cite{pace}\begin{equation}\label{x15}
\bigtriangledown ~.~\Big[(\vec{u}.\bigtriangledown)\vec{u}\Big]=\frac{1}{3}\theta^{2}+\sigma^{2}-w^{2},
\end{equation}
where 
\begin{equation}\label{x16}
\theta=\bigtriangledown_{\vec{x}}~.~\vec{u}~,
\end{equation}
\begin{equation}\label{x17}
\sigma_{ij}=\frac{1}{2}\Big(\frac{\partial u^{j}}{\partial x^{i}}+\frac{\partial u^{i}}{\partial x^{j}}\Big)-\frac{1}{3}\theta \delta_{ij}~,
\end{equation}
\begin{equation}\label{x18}
w_{ij}=\frac{1}{2}\Big(\frac{\partial u^{j}}{\partial x^{i}}-\frac{\partial u^{i}}{\partial x^{j}}\Big),
\end{equation}
Here, $\sigma^{2}=\sigma_{ij}\sigma^{ij}$ is the shear tensor and $w^{2}=w_{ij}w^{ij}$ is the rotation tensor.
Taking time derivative of Eq. (\ref{x12}) and using Eqs. (\ref{x12}), (\ref{x13}), (\ref{x14}) and (\ref{x15}), one can obtain \cite{pace}
\begin{eqnarray}\label{x19}
\ddot{\delta}+\Big(2H-\frac{\dot{\omega}}{1+\omega}\Big)\dot{\delta}-\frac{4+3\omega}{3(1+\omega)}
\frac{\dot{\delta}^{2}}{1+\delta}-4\pi G \bar{\rho}(1+\omega)(1+3\omega)\delta (1+\delta)-
(1+\omega)(1+\delta)(\sigma^{2}-w^{2})=0.
\end{eqnarray}
Now, we can introduce  the following definition 
\begin{equation}\label{cvbn}
\frac{\partial}{\partial t}=a H(a) \frac{\partial}{\partial a}.
\end{equation}
Using Eq. (\ref{cvbn}) and inserting Eq. (\ref{x19}), one can rewrite Eq. (\ref{x19}) as follows \cite{pace}\begin{eqnarray}\label{x20}
\delta^{''}+\left(\frac{3}{a}+\frac{E^{'}}{E}-\frac{\omega^{'}}{1+\omega}\right)\delta^{'}-\frac{4+3\omega}{3(1+\omega)}
\frac{{\delta^{'}}^{2}}{1+\delta}-\frac{3}{2}\frac{\Omega_{fluid,0}}{a^{2}E^{2}(a)}h(a)(1+\omega)(1+3\omega)\delta (1+\delta)-
~~~~~~~~~~~~~~~~~~~~~~\nonumber\\
\frac{1}{aH^{2}(a)}(1+\omega)(1+\delta)(\sigma^{2}-w^{2})=0,~~~~~~~~~~~~~~~~~~~~~~~
\end{eqnarray} 
where $E=H/H_{0}$ is the dimensionless Hubble parameter, $\Omega_{fluid,0}=8\pi G \bar{\rho}/3H^{2}$ is the density parameter of the fluid at 
$a_{0}=1$, $h(a)$ is a function that describes the time evolution of  dark energy
with scale factor, $H_{0}$ is  the Hubble parameter  at the present time and
the prime sign denotes  the derivative with respect to the scale factor. 

For the collapse of a homogeneous sphere, one can ignore the shear and rotation tensors. Also, we limit ourselves to the spherical perturbation filled with
dust $\omega=0$, for which
$$h(a)=a^{-3}.$$
Therefore, one can obtain the non-linear and linear perturbation equations as follows \cite{pace}, \cite{malekjani51}\begin{equation}\label{x21}
\delta^{''}+\left(\frac{3}{a}+\frac{E^{'}}{E}\right)\delta^{'}-\frac{4}{3}
\frac{{\delta^{'}}^{2}}{1+\delta}-\frac{3}{2}\frac{\Omega_{m_{0}}}{a^{5}E^{2}(a)}\delta (1+\delta)=0,
\end{equation}
\begin{equation}\label{x22}
\delta^{''}+\left(\frac{3}{a}+\frac{E^{'}}{E}\right)\delta^{'}-\frac{3}{2}\frac{\Omega_{m_{0}}}{a^{5}E^{2}(a)}\delta=0.
\end{equation}

\section*{\textbf{\textit{Appendix II}}}

In the numerical study of solving Eqs.  (\ref{delta}) and (\ref{delta1}), we have used the procedure used in Ref.\cite{pace} 
to obtain the initial conditions for drawing the curves of figure 3, in tachyon scalar field model, as follows
\begin{equation}\label{app6}
g(a_{i})=3.615\times 10^{-4},~~~~~~~~~~g^{'}(a_{i})=0,~~~~~~~~~~~a_{i}=10^{-4}.
\end{equation}
The scale factor $a$ at the redshift $z$ is defined as follows\begin{equation}\label{app2}
a=(1+z)^{-1}.
\end{equation}
Taking derivative of Eq. (\ref{app2}), one can obtain
\begin{equation}\label{app3}
\frac{d}{dz}=-a^{2}\frac{d}{da}.
\end{equation} 
Using Eqs. (\ref{eos}), (\ref{omegalam}) and  (\ref{app3}) and inserting Eqs.  (\ref{df}), (\ref{ddf}), we can obtain 
\begin{equation}\label{app4}
\frac{d\Omega_{\Lambda}}{da}+\frac{3\Omega_{\Lambda}}{a}(1-\Omega_{\Lambda})(\frac{2}{3n}-1)=0,
\end{equation}
\begin{equation}\label{app5}
\frac{dE}{da}+\frac{3}{2}\frac{E}{a}\Big[1+\Omega_{\Lambda}(\frac{2}{3n}-1)\Big]=0.
\end{equation}
Using Eqs. (\ref{app4}), (\ref{app5}), (\ref{app6}) and inserting Eq. (\ref{delta1}), we can plot the curve of figure $(3-a)$ for tachyon scalar field model 
 using the  ``Mathematica" in the linear case. Also, using Eqs. (\ref{friedman}), (\ref{Omega}), (\ref{Omega5}) and $\rho_{m}=\rho_{m_{0}}a^{-3}$ and $\rho_{\Lambda}=\rho_{\Lambda_{0}}a^{-3(1+\omega_{\Lambda})}$, one can obtain
\begin{equation}\label{app7}
E(a)=\sqrt{\Omega_{m_{0}}a^{-3}+\Omega_{\Lambda_{0}}a^{-3(1+\omega_{\Lambda})}}.
\end{equation}
Also, using Eqs. (\ref{omegalam}), (\ref{app6}), (\ref{app7}) and inserting Eq. (\ref{delta}), we can plot the curve of figure $(3-b)$ for tachyon scalar field model 
using the ``Mathematica" in the non-linear case.

\section*{Acknowledgment}
We would like to thank M. Malekjani  for giving us
useful comments that helped us to improve the scientific content of the manuscript.

\label{lastpage}


\begin{thebibliography}{}
\bibitem[\protect\citeauthoryear{ Abramo}{2007}]{abramo} Abramo.  L. R., Batista. R. C., Liberato. L., Rosenfeld. R., 2007, JCAPP, 11, 12.

\bibitem[\protect\citeauthoryear{Avelino}{2011}]{losano} Avelino. P. P, Losano. L, Rodrigues. J. J, 2011, Phys. Lett. B 699, 10.

\bibitem[\protect\citeauthoryear{Bond}{1991}]{bond}  Bond. R. J.,  Cole.
S.,  Efstathiou. G.,  Kaiser. N. 1991, ApJ, 379 ,440.  
\bibitem[\protect\citeauthoryear{Bernardeau}{1994}]{bernard}  Bernardeau.
F. 1994, ApJ, 433, 1.
\bibitem[\protect\citeauthoryear{Bernardis}{2000}]{bernardis} 
 Bernardis. P. De., et al. 2000, Nature 404, 955;\\
 Perlmutter. S., et al. 2003, AStrophys. J. 598, 102;\\
 Seljak. U., et al. 2005, Phys. Rev. D 71, 103515. 
\bibitem[\protect\citeauthoryear{Cai}{2010}]{msetare}  Cai.  Y. F., Saridakis.
 E. N., Setare.  M. R., Xia. J. Q., 2010, 
Phys. Rept. 493, 1. 
\bibitem[\protect\citeauthoryear{Copeland}{2006}]{copeland}  Copeland. E.
J., Sami. M.,  Tsujikawa. T. S. 2006, IJMPD, 15, 1753.  
\bibitem[\protect\citeauthoryear{Doran}{2006}]{doran}  
 Doran. M., Robbers. G. 2006, JCAPP, 6, 26;\\ Wetterich. C. 2004, Phys. Lett. B, 594, 17. 
\bibitem[\protect\citeauthoryear{Fillmore}{1984}]{fill} Fillmore.  A. J.,  Goldreich. P. 1984, ApJ, 281, 1.  
\bibitem[\protect\citeauthoryear{Gunn}{1972}]{gun} Gunn.  J. E., Gott. R. J. 1972, ApJ, 176,1. 
\bibitem[\protect\citeauthoryear{Hoffman}{1985}]{haffman}  Hoffman. Y., Shaham.
 J. 1985, ApJ. 297, 16.  
\bibitem[\protect\citeauthoryear{Meyer}{2012}]
{meyer} Meyer.  S.,  Pace. F., Bartelmann.  M. 2012, Phys.Rev.D, 86, 103002.
\bibitem[\protect\citeauthoryear{Pace}{2010}]
{pace} Pace.  F.,  Waizmann. J. C., Bartelmann.  M., 2010, MNRAS, 406, 1865.
\bibitem[\protect\citeauthoryear{Pace}{2014}]
{pace1}  Pace. F., Moscardini.  L,  Crittenden. R.,  Bartelmann. M.,
 Pettorino. V. 2014, MNRAS, 437, 547.
 \bibitem[\protect\citeauthoryear{Padmanabhan}{1996}]{padmanabhan} Padmanabhan.
T. 1996, Cosmology and Astrophysics through Problems, Cambridge University
Press.
\bibitem[\protect\citeauthoryear{Padmanabhan}{2003}]{padmanabhan1}  Padmanabhan. T. 2003, Phys. Rept. 380, 235. 
\bibitem[\protect\citeauthoryear{Press}{1974}] {press}  Press. H. W.,  Schechter.
P. 1974, ApJ, 187, 425. 
\bibitem[\protect\citeauthoryear{Percival}{2005}]
{percival} Percival.  W. J. 2005, A. A, 443, 819. 
\bibitem[\protect\citeauthoryear{Popolo}{2006}] {del}  Del Popolo.  A, 2006, ApJ, 637, 12;\\
Del Popolo.  A, 2006, A. A, 448, 439.
\bibitem[\protect\citeauthoryear{Riess}{1998}]{Riess} Riess.  A. G., et al. 1998, Astron. J. 116, 1009;\\
 Perlmutter. S., et al. 1999, AStrophys. J. 517, 565. 
\bibitem[\protect\citeauthoryear{Ryden}{1987}]{ryden}  Ryden. S. B., Gunn.
E. J. 1987, ApJ. 318, 15.
\bibitem[\protect\citeauthoryear{Setare}{2007}]{setare}  Setare.  M. R. 2007, Phys. Lett. B 653,116;\\ Setare. M. R., Sadeghi. J., Amani.  A. R., 2009, Phys. Lett. B 673, 241.
\bibitem[\protect\citeauthoryear{Sheth}{1999}] {sherth}  Sheth. R. K., Tormen.
G., 1999, MNRAS, 308, 119;\\
 Sheth. R. K., Tormen. G. 2002, MNRAS, 329, 61.
\bibitem[\protect\citeauthoryear{Viana}{1996}] {viana}  Viana. P. T. P.,  Liddle. A. R., 1996, MNRAS, 281, 323.
\bibitem[\protect\citeauthoryear{Wang}{1998}]{ws} Wang.  L., Steinhardt.
 P. J. 1998, Astro. Phys. J. 508, 483.  
\bibitem[\protect\citeauthoryear{Weinberg}{1989}]{weinberg} 
 Weinberg. S. 1989, Reviews of Modern Physics, 61, 1.
\bibitem[\protect\citeauthoryear{Yahagi}{2004}] {yahagi} Yahagi. H., Nagashima.
M., Yoshii. Y. 2004, ApJ, 605, 709.
\bibitem{padd}  Padmanabhan. T, 2002, Phys. Rev. D 66, 021301.
\bibitem{malekjani}  Malekjani. M,  Naderi. T, Pace. F, 2015, MNRAS 453, 4148.
\bibitem{holog}  Naderi. T,  Malekjani. M, Pace. F, 2015, MNRAS, 447, 1873.

\bibitem[\protect\citeauthoryear{Zhang}{2007}]{Zhang.}  Zhang. J, Zhang. X, Liu. H, 2007,Phys. Lett. B 651, 84.

\bibitem{malekjani51}  Malekjani. M.,  Lu. J,  Nazari-Pooya. N.,  Xu. L,  Mohammad-Zadeh Jassur. D.,  Honari-Jafarpour.  M., 2015, 
Astrophys Space Sci, 360, 24.

\end{thebibliography}
\end{document}